\documentclass[twocolumn,superscriptaddress,amsmath,amssymb,prb,longbibliography]{revtex4-2}
\usepackage{graphicx}
\usepackage{epsfig}
\usepackage{epstopdf}
\DeclareMathAlphabet\mathbfcal{OMS}{cmsy}{b}{n}
\usepackage{relsize}
\usepackage{amsmath}
\usepackage{amssymb}
\usepackage{ulem}
\usepackage{color}
\begin{document}
\title{Ultrafast valley polarization in bilayer graphene}
\author{Pardeep Kumar}
\email[]{mehra.pardeep89@gmail.com}
\affiliation{Center for Nano-Optics (CeNO) and Department of Physics and Astronomy,  Georgia State University, Atlanta, Georgia 30303, USA}

\author{Thakshila M. Herath}
\affiliation{Center for Nano-Optics (CeNO) and Department of Physics and Astronomy,  Georgia State University, Atlanta, Georgia 30303, USA}

\author{Vadym Apalkov}
\affiliation{Center for Nano-Optics (CeNO) and Department of Physics and Astronomy,  Georgia State University, Atlanta, Georgia 30303, USA}


\date{\today}

\begin{abstract}
We study theoretically interaction of a bilayer graphene with a circularly polarized ultrafast optical pulse of a single oscillation at an oblique incidence. The normal component of the pulse breaks the inversion symmetry of the system and opens up a dynamical band-gap, due to which a  valley-selective population of the conduction band becomes sensitive to the angle of incident of the pulse. We show that the magnitude of the valley polarization can be controlled by the angle of incidence,  the amplitude,  and  the angle of in-plane polarization of the chiral optical pulse. Subsequently, a sequence of a circularly polarized pulse followed by a linearly polarized femtosecond-long pulse can be used to control the valley polarization created by the preceding pulse.  Generally, the linearly polarized pulse depolarizes the system. The magnitude of such a depolarization depends on the amplitude,  and the in-plane polarization angle of the linearly polarized pulse.  Our protocol provides a favorable platform for applications in valleytronics.   
\end{abstract}


\maketitle

\section{Introduction}
In recent years, the interaction of optical fields with different systems such as vapor phase media \cite{lvosky2009}, mechanical \cite{kumar2019} and solid state materials \cite{sun2018} have been explored extensively for quantum information processing. Such light-matter interaction can be used to control different degrees of freedom such as spin \cite{igor2015}, charge \cite{hollenberg2004}, mechanical \cite{wang2012}, and rotational \cite{shi2016} for storage and processing of information. Interestingly, the existence and manipulation of valley  degree of freedom \cite{rycerz2007} in crystalline solids has provided a remarkable platform for future electronics known as \textit{valleytroincs} \cite{schaibley2016}. The pivotal idea is to create imbalance in populations of two valleys thereby giving rise to valley polarization \cite{gunlycke2011}. Recently, the valley polarization has been manipulated in 2D materials by means of strain \cite{low2010}, electric \cite{liu2014}, magnetic \cite{xiao2007}, and optical \cite{yao2008} fields. Specifically, 2D materials with honeycomb lattice structures, such as graphene \cite{neto2009} and transition metal dichalcogenides \cite{liu2015} (TMDC), are of particular interest due to existence of inequivalent valleys at the $K$ and $K^{\prime}$ points of the Brillouin zone. Furthermore, if the inversion symmetry in such systems is broken, which happens in  TMDC, gapped graphene \cite{pedersen2009}, and biased bilayer graphene (BLG) \cite{mccann2013,rozhkova2016}, then the two valleys respond differently to circularly polarized light, thereby resulting in selective population of one of the valleys and eventually to the valley polarization \cite{cao2012}.

The emergence of the ultrafast laser technology paves a versatile pathway to coherent control of electron dynamics at a femtosecond  time scale \cite{krausz2014}. Moreover, strong ultrafast optical pulses provide a favorable platform to probe extremely nonlinear behavior of 2D materials \cite{ghimire2014}. The interaction of such optical pulses with graphene causes dramatic changes in its electron dynamics \cite{heide2018} and produces valley currents \cite{higuchi2017}. For example, a single oscillation of a linearly polarized pulse results in quantum interference due to a double passage by an electron of the Dirac points in graphene \cite{kelardeh2015}. Consequently, the conduction band (CB) population distribution in the reciprocal space of graphene shows interference fringes. A circularly polarized pulse of a few optical cycles, on the other hand, equally  populates the $K$ and $K^{\prime}$ valleys \cite{kelardeh2016} in graphene, thanks to the presence of inversion symmetry. However, such chiral optical pulses play an important role in honeycomb lattices with broken inversion symmetry, where they selectively populate $K$ or $K^{\prime}$ valley, which ultimately leads to the finite valley polarization \cite{mak2012,zeng2012}. Recently, an extremely high valley polarization has been predicted in gapped graphene \cite{azar2018} and TMDC \cite{azar20182} placed in the field of a chiral optical pulse of a single oscillation. Such high valley polarization is attributed to the topological resonance, which arises due to the mutual cancellation of the dynamic and geometric phases.  
  
Contrary to monolayer,  BLG \cite{mccann2013,rozhkova2016} is of particular interest due to its enhanced electrical and thermal properties \cite{castro2007,gong2012} and specific optical signatures \cite{yan2012}.  The BLG can exist in three configurations: (i) AA stacking, when each carbon atom in one layer overlaps with the corresponding atom of another layer, (ii) AB stacking (Bernal stacking), when one of the carbon atoms $B_{b}$ of the bottom layer  is placed exactly below the carbon atom $A_{t}$ of the top layer, as shown in Fig. \ref{fig1}(a), and (iii) twisted bilayer, the configuration in which the top graphene layer is rotated by a small angle with respect to the bottom layer \cite{bistritzer2011}.  The existence of parabolic energy dispersion \cite{mccann2006}, massive chiral quasiparticles \cite{li2007}, Berry phase \cite{novoselov2006} of 2$\pi$ and the flexibility to tune each individual layer \cite{ohta2006, mccann2006_v2,oostinga2007} in BLG makes it potentially different from its monolayer \cite{geim2007,liu2011,xiao2010,mikitik1999,novoselov2005,baringhaus2014,young2009} counterpart.  Similar to gapped graphene and TMDC, the inversion symmetry \cite{zhang2010} in BLG can be also  removed by introducing a difference in the on-site energies by a perpendicular electric field.  As a result, the low-energy bands exhibit a bandgap at the $K$ and $K^{\prime}$ points. The bandgap can be tuned by the transverse electric field.  It is to be noted that, unlike monolayer graphene, the existence of pairs of conduction bands (CBs) and valence bands (VBs) in BLG, makes it  a suitable candidate for applications in valleytronics. This is because in BLG one can selectively populate one of the CBs by tuning the excitation energy and hence control the population of different CBs in different valleys.  This provides an extra opportunity to manipulate the corresponding valley polarization \cite{mak2012} in BLG.

In this paper, we study the interaction of a single oscillation chiral optical pulse with AB-stacked BLG. The circularly polarized pulse is applied at oblique incidence, so that the normal component of the optical field brakes the inversion symmetry and opens a dynamical band gap at the $K$ and $K^{\prime}$ points. This gives rise to a valley polarization which can be manipulated by the angle of incidence, the amplitude and the in-plane orientation of the circular pulse.  We also consider a sequence of a circularly polarized pulse followed by a \textit{normally} incident femtosecond-long linearly polarized pulse that changes the valley polarization generated by the circularly polarized pulse in BLG. Generally, the linearly polarized pulse depolarizes the system. Here we show that the magnitude of such a depolarization depends on the amplitude, and the angle of polarization of the linearly polarized pulse. It is to be noted that here the linearly polarized pulse is incident normally to BLG and itself does not give rise to any inversion symmetry breaking. This is unlike a situation in \cite{kumar2020}, where linearly polarized pulse is incident at an angle and its normal component opens a band gap at the Dirac points due to which a current is generated along the symmetry axis of BLG for perpendicular polarization of the linearly polarized pulse. Moreover, in the present paper, we employ a sequence of a chiral pulse at oblique incident (which simultaneously breaks the inversion and time-reversal symmetry) followed by a linearly polarized pulse which solely manoeuvre the valley polarization generated by the former circular pulse.

The paper is organized as  follows. In Sec. \ref{model} we describe the model of BLG and introduce the main equations. In Sec. \ref{results} we present and discuss our main results. Finally, concluding remarks are presented in Sec. \ref{conclusion}.

\section{Model}
\label{model}
\begin{figure}[ht!]
	\begin{center}
		\begin{tabular}{cc}
			\includegraphics[scale=1.05]{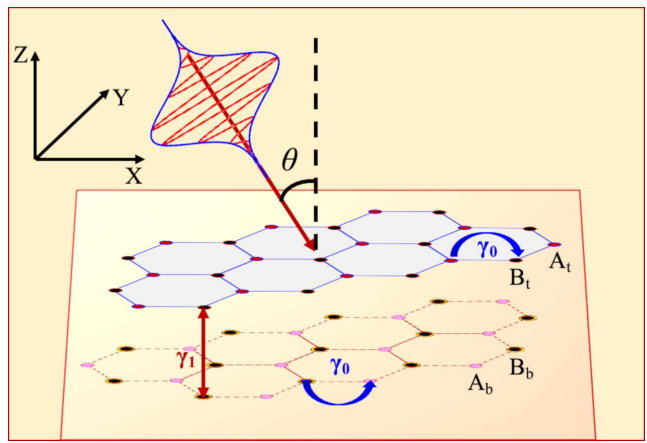}
		\end{tabular}
		\caption{ BLG in Bernal (AB) stacking, where the honeycomb lattice structure of two graphene monolayers are coupled in such a way that each carbon atom in  sublattice $A_{t}$ of top layer is exactly above the carbon atom in sublattice $B_{b}$ of the bottom layer. The bottom (top) layer is represented by dashed (solid) red (blue) lines.  The BLG interacts with a chiral optical pulse at an oblique incidence $\theta$ and in-plane orientation that is determined by angle $\phi$. }
		\label{fig1}
	\end{center}
\end{figure}

We consider AB-stacked BLG whose tight-binding Hamiltonian is given by
\begin{align}
H_{0}=\begin{bmatrix}
0 & -\gamma_{0} f\left(\mathbf{k}\right)& 0 & 0\\
-\gamma_{0} f^{\ast}\left(\mathbf{k}\right) & 0 & \gamma_{1} & 0\\
0 & \gamma_{1} & 0 & -\gamma_{0} f\left(\mathbf{k}\right)\\
0 & 0 & -\gamma_{0} f^{\ast}\left(\mathbf{k}\right) & 0
\end{bmatrix}\;,\label{eq1}
\end{align}
where $\gamma_{0}=3.16$ eV is the intralayer hopping integral, $\gamma_{1}=0.381$ eV is the interlayer hopping integral, which represents the coupling between the orbitals of the dimer sites, and $f(\mathbf{k})$ is given by the following expression
\begin{align}
f(\mathbf{k})=\exp\left(\frac{iak_{y}}{\sqrt{3}}\right)+2\exp\left(-\frac{iak_{y}}{2\sqrt{3}}\right)\cos\left(\frac{ak_{x}}{2}\right)\;,\label{eq2}
\end{align}
where $a=2.46~\mbox{\AA}$ is the lattice constant. 

\begin{figure}[ht!]
\begin{center}
\begin{tabular}{c}
\includegraphics[scale=0.45]{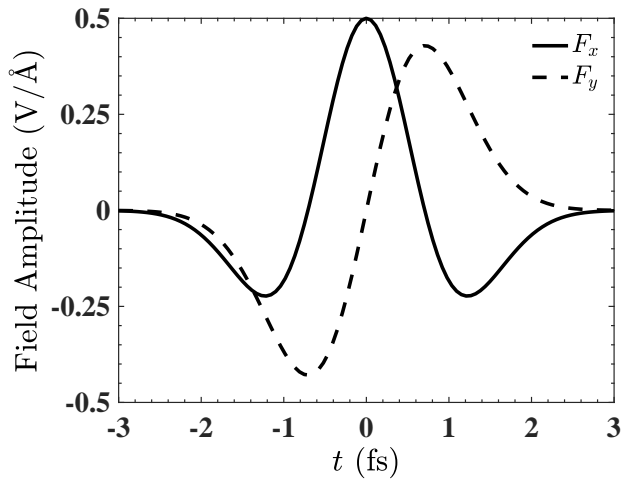}
\end{tabular}
\caption{The pulse profile as a function of time.  The parameters of the pulse are $F_{0}=0.5$ V/$\mbox{\AA}$, $\tau=1$ fs, $\phi=0^{\circ}$, and $\theta=0^{\circ}$.}
\label{fig2}
\end{center}
\end{figure}

The AB-stacked BLG interacts with a right-handed ultrafast circularly polarized of single oscillation,  as shown in Fig. \ref{fig1}.  This chiral pulse is incident at an angle $\theta$ such that its perpendicular component breaks the inversion symmetry and opens a band-gap at the Dirac points.  Further, the in-plane component of the circular pulse can be manipulated both by angle of incident $\theta$ and in-plane orientation $\phi$.   Since the duration of the pulse is less than the characteristic electron scattering time of 10-100 fs  \cite{breusing2011,malic2011,hwang2008}, the electron dynamics in the presence of the optical pulse is coherent and can be described by the time-dependent Schr\"{o}dinger equation
\begin{align}
i\hbar\frac{d\Psi}{dt}=H(t)\Psi\;.\label{eq3}
\end{align}  
where
\begin{align}
H(t)=H_{0}-e\mathbf{F_{p}}(t)\mathbf{r}-\frac{eL_{z}F_{z}(t)}{2}\begin{bmatrix}
1 & 0& 0 & 0\\
0 & 1 & 0 & 0\\
0 & 0 & -1 & 0\\
0 & 0 & 0 & -1\end{bmatrix}\;,\label{eq4}
\end{align} 
where  $\mathbf{F_{p}}(t)=(F_{x}\cos\theta\cos\phi+F_{y}\sin\phi,-F_{x}\cos\theta\sin\phi+F_{y}\cos\phi)$ is the in-plane component of the electric field of the pulse and $F_{z}(t)=F(t)\sin\theta$. Here $F_z(t)$ component breaks the interlayer symmetry of the system and opens a dynamic band gap at the $K$ and $K^{\prime}$ points. On the other hand,  the form $\mathbf{F_{p}}(t)$ suggests that both angle of incidence and the angle of polarization control its in-plane  orientation thereby providing ellipticity to the circular pulse.  The applied ultrafast circularly polarized pulse is shown in  Fig. \ref{fig2} and is parametrized by the following equations 
\begin{align}
F_{x}&=F_{0}(1-2u^{2})e^{-u^{2}}\;,\label{eq5}\\
F_{y}&=2F_{0}ue^{-u^{2}}\;,\label{eq6}
\end{align}
where $F_{0}$ is the amplitude of the optical pulse, and $u=t/\tau$, where $\tau$ is the pulse duration, which we consider to be 1 fs. 

Such ultrafast optical pulses of single oscillation gives rise to both interband and intraband electron dynamics. For BLG a detailed description of such electron dynamics and governing equations,  we follow the complete methodology presented in  \cite{kumar2020}. 

\section{Results}
\label{results}
The band structure of BLG  can be computed from Eq. (\ref{eq1}) and consists of two valence ($V_{1},V_{2} $) and two conduction bands ($C_{1},C_{2}$). We solve the electron dynamics using the initial conditions that the valence bands 1 and 2 are occupied \cite{fatemeh2018}, i.e. $(\beta_{C_{2}\mathbf{q}},\beta_{C_{1}\mathbf{q}},\beta_{V_{1}\mathbf{q}},\beta_{V_{2}\mathbf{q}})=(0,0,1,0)$ and $(\beta_{C_{2}\mathbf{q}},\beta_{C_{1}\mathbf{q}},\beta_{V_{1}\mathbf{q}},\beta_{V_{2}\mathbf{q}})=(0,0,0,1)$,. Here, $\beta_{\alpha\mathbf{q}}~(\alpha=V_{1},V_{2},C_{1},C_{2})$ are the expansion coefficients whose detail is presented in \cite{kumar2020}. From this solution we calculate the residual CB populations, i.e., the CB populations after the pulse, $N_{C_{2}}^{(\mbox{res})}(\mathbf{q})=|\beta_{C_{2}\mathbf{q}}(t=\infty)|^{2}$ and $N_{C_{1}}^{(\mbox{res})}(\mathbf{q})=|\beta_{C_{1}\mathbf{q}}(t=\infty)|^{2}$. 

\subsection{Circularly polarized pulse}

\begin{figure}[ht!]
	\begin{center}
		\begin{tabular}{cc}
			\includegraphics[scale=0.8]{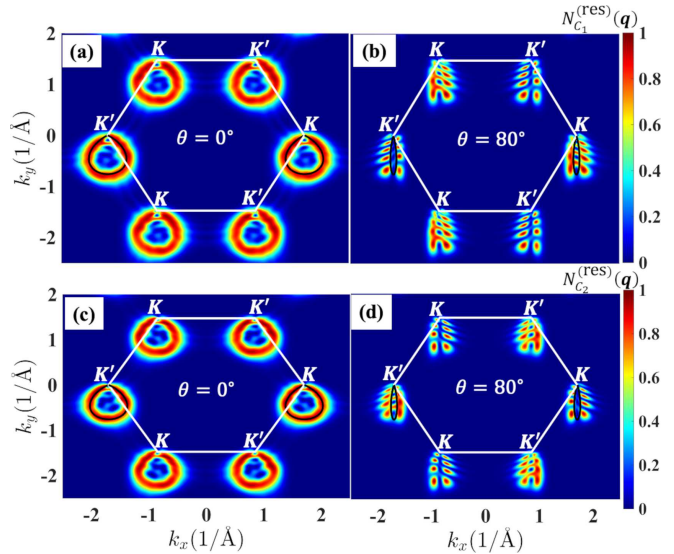} 
		\end{tabular}
	\end{center}
	\caption{Residual CB population distribution (a,b) $N^{\mbox{(res)}}_{C_{1}}(\mathbf{k})$ and (c,d) $N^{\mbox{(res)}}_{C_{2}}(\mathbf{k})$ at the end of right-handed circularly polarized pulse for the first and the second CBs of BLG for different values of the angle of incidence ($\theta$). The boundary of the first Brillouin zone is shown by white lines and separatrix is shown by black lines. The amplitude of the pulse is fixed at $F_{0}=0.5$ V/$\mbox{\AA}$, $\phi=0^{\circ}$, and the rest of parameters are the same as in Fig. \ref{fig2}.}
	\label{fig3}
\end{figure}

We consider a right-handed circularly polarized pulse. The residual CB population distribution  
for CBs $C_{1}$ and $C_{2}$ is shown in Fig. \ref{fig3} in the first Brillouin zone for different angles of incidence and fixed amplitude of the pulse, $F_0 =0.5$ V/$\mbox{\AA}$. Here, the black lines within the bright colored region of CB population represents the separatrix which is defined as the set of initial conditions for which the electron trajectories pass exactly through the Dirac points and is governed by the following equation

\begin{align}
\mathbf{q}(t)&=\mathbf{K}-\mathbf{k}(0,t), ~~~\mbox{or}~~~\mathbf{q}(t)=\mathbf{K^{\prime}}-\mathbf{k}(0,t)\;,\label{eq7}
\end{align}

As depicted in Fig.\ref{fig3}, for the normal incidence, $\theta = 0^{\circ}$, the residual CB population is concentrated near the separatrix. It is symmetric with respect to the $k_y$-axis for both $C_{1}$ and $C_{2}$ CBs. In the final state, the system is valley unpolarized, i.e., the total CB populations of the $K$ and $K^{\prime}$ valleys are the same, see Fig. \ref{fig3}(a,c).  However, if the angle of incidence is $\theta=80^{\circ}$, the population distribution becomes asymmetric with respect to the $k_y$-axis, the  populations of the $K$ and $K^\prime$ valleys become different, and the BLG acquires a valley polarization [see Fig. \ref{fig3}(b,d)]. It is to be noted that for normal incidence the pulse remains circular but due to the absence of inversion symmetry it does not create valley polarization. However, oblique incidence not only results   a normal component which breaks the inversion symmetry but also produces ellipticity of the incident pulse. This is why the separatrix becomes elliptical in Fig. \ref{fig3}. 

\begin{figure}[ht!]
	\begin{center}
		\begin{tabular}{cc}
			\includegraphics[scale=0.4]{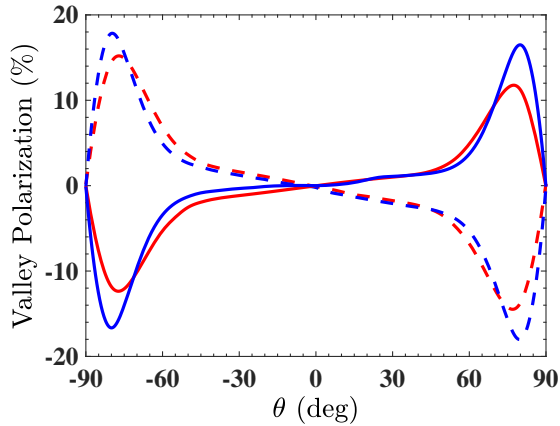} 
		\end{tabular}
	\end{center}
\caption{Valley polarization after circularly polarized pulse for the first CB, $\eta_{C_{1}}$ (solid lines) and the second CB, $\eta_{C_{2}}$ (dashed lines),   as a function of angle of incidence of the circularly polarized pulse.  Here the amplitude of the pulse is  $F_{0}=0.3$ V/$\mbox{\AA}$ (red lines) and  $F_{0}=0.5$ V/$\mbox{\AA}$ (blue lines). Other parameters are same as in Fig. \ref{fig3}.}
	\label{fig4}
\end{figure}

As shown above,  the angle of incidence of the circularly polarized pulse plays an important role to manipulate the  CB population distributions around Dirac points thereby producing valley polarization of the system. To further quantify this effect,  we define the valley polarization \cite{schaibley2016,mak2012,zeng2012} for CBs $C_{1}$ and $C_{2}$ as follows:
\begin{align}
\eta_{i}=\frac{n_{i}^{K}-n_{i}^{K^{\prime}}}{n_{i}^{K}+n_{i}^{K^{\prime}}}~~~~~~~~(i=C_{1},C_{2})\;,\label{eq8}
\end{align} 
where $n_{i}^{K}=\sum_{k\in K}N^{\mbox{(res)}}_{i}(\mathbf{k})$ is the total CB population of the $K$ valley and $n_{i}^{K^{\prime}}$ has the same definition for $K^{\prime}$-valley.    The residual valley polarization generated by the circular pulse depends strongly on the angle of incidence, as shown in Fig. \ref{fig4}. Clearly, for normal incidence ($\theta=0^{\circ}$) no valley polarization exists. This is because for $\theta=0^{\circ}$ there is no inversion symmetry breaking and as a result the residual CB population around Dirac points remain symmetric [Fig.  \ref{fig3}(a,c)] thereby producing zero valley polarization.  However,  for oblique incidence ($\theta \neq 0^{\circ}$), the normal component of the pulse breaks the inversion symmetry and opens a band-gap at the Dirac points. This results in asymmteric residual CB population  $K$ and $K^{\prime}$-points and generates valley polarization, as shown in Fig. \ref{fig4}.  For $\theta> 0^{\circ}$,  the residual CB population for $C_{1}~(C_{2})$ at $K~(K^{\prime})$ valley dominates over the $K^{\prime}~(K)$ valley. Consequently, the valley polarization, $\eta_{C_{1}}~(\eta_{C_{2}})$ is positive (negative) and attains a maximum value of about $\sim 16\% (-18\%)$ for $\theta=80^{\circ}$.  The effect is reversed for the negative values of $\theta$, as shown in Fig. \ref{fig4}.  It is to be noted that the fundamental difference between the monolayer and BLG is that, only in BLG the perpendicular component of the 
electric field of the pulse opens a dynamic bandgap, which favors the valley polarization. As a result, the valley polarization of BLG depends on the angle of incidence, $\theta $, of the pulse, but there is  no such dependence for monolayer graphene. 
\begin{figure}[ht!]
	\begin{center}
		\begin{tabular}{cc}
			\includegraphics[scale=0.4]{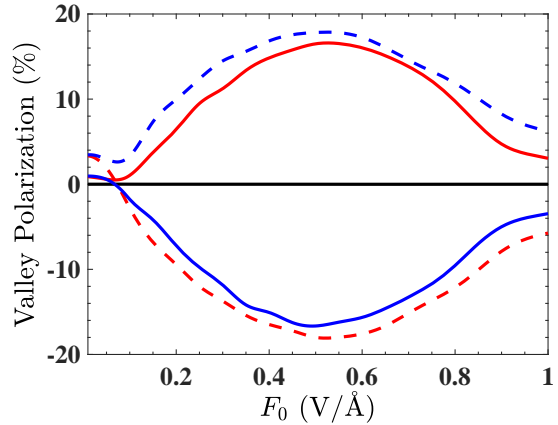} 
		\end{tabular}
	\end{center}
\caption{Valley polarization after circularly polarized pulse for the first CB, $\eta_{C_{1}}$ (red lines) and the second CB, $\eta_{C_{2}}$ (blue lines),   as a function of amplitude of the circularly polarized pulse.  Here, $\theta=80^{\circ}$ (solid lines), $\theta=-80^{\circ}$ (dashed lines) and $\phi=0^{\circ}$.  Other parameters are same as in Fig. \ref{fig3}.}
\label{fig5}
\end{figure}

Now,  to further describe the valley polarization in BLG, we fix $\theta=\pm 80^{\circ}$. Apart from angle of incidence, the residual CB population around the Dirac points and hence the valley polarization can be manipulated by the amplitude of the applied pulse. Such a dependence of the valley polarization on  the field amplitude is shown in Fig. \ref{fig5}. For $\theta=80^{\circ}$, and at a given field amplitude, residual CB population for $C_{1}$ around $K$ valley dominates over the $K^{\prime}$ valley. This gives rise to a positive $\eta_{C_{1}}$ for different field amplitudes and it attains a maximum value of $\sim 16 \%$ at $F_0 = 0.5~\mbox{V}$/\AA. On the other hand,  $\eta_{C_{2}}$ remains negative for strong field amplitudes and it achieve a highest negative value of $\sim -18\%$ at $F_{0}=0.5~\mbox{V}/\mbox{\AA}$.  Further, both $\eta_{C_{1}}$ and $\eta_{C_{2}}$ not only switches sign but also becomes larger  for $\theta=-80^{\circ}$, as depicted in Fig. \ref{fig5}. 

\begin{figure}[ht!]
	\begin{center}
		\begin{tabular}{cc}
			\includegraphics[scale=0.4]{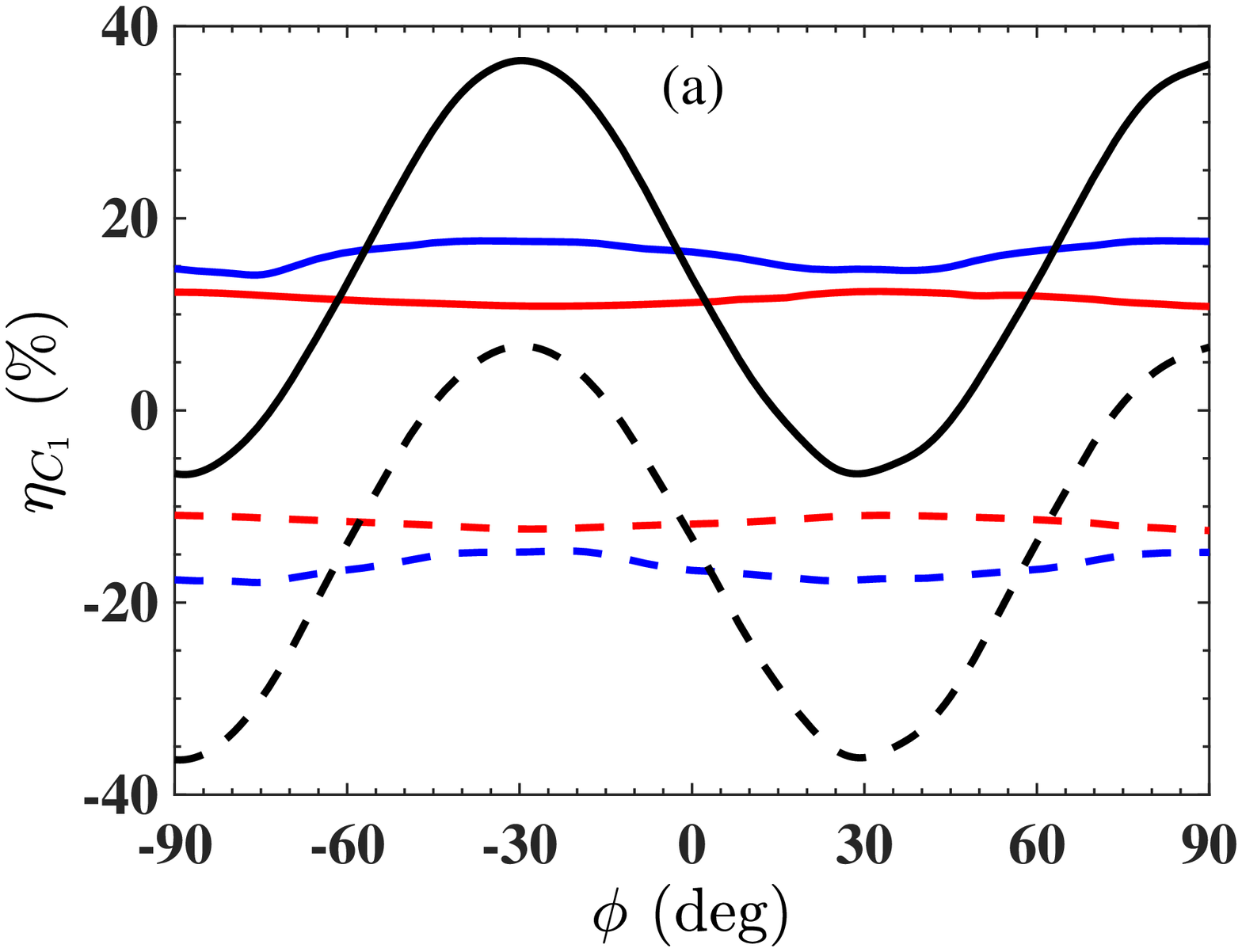}\\
			\includegraphics[scale=0.4]{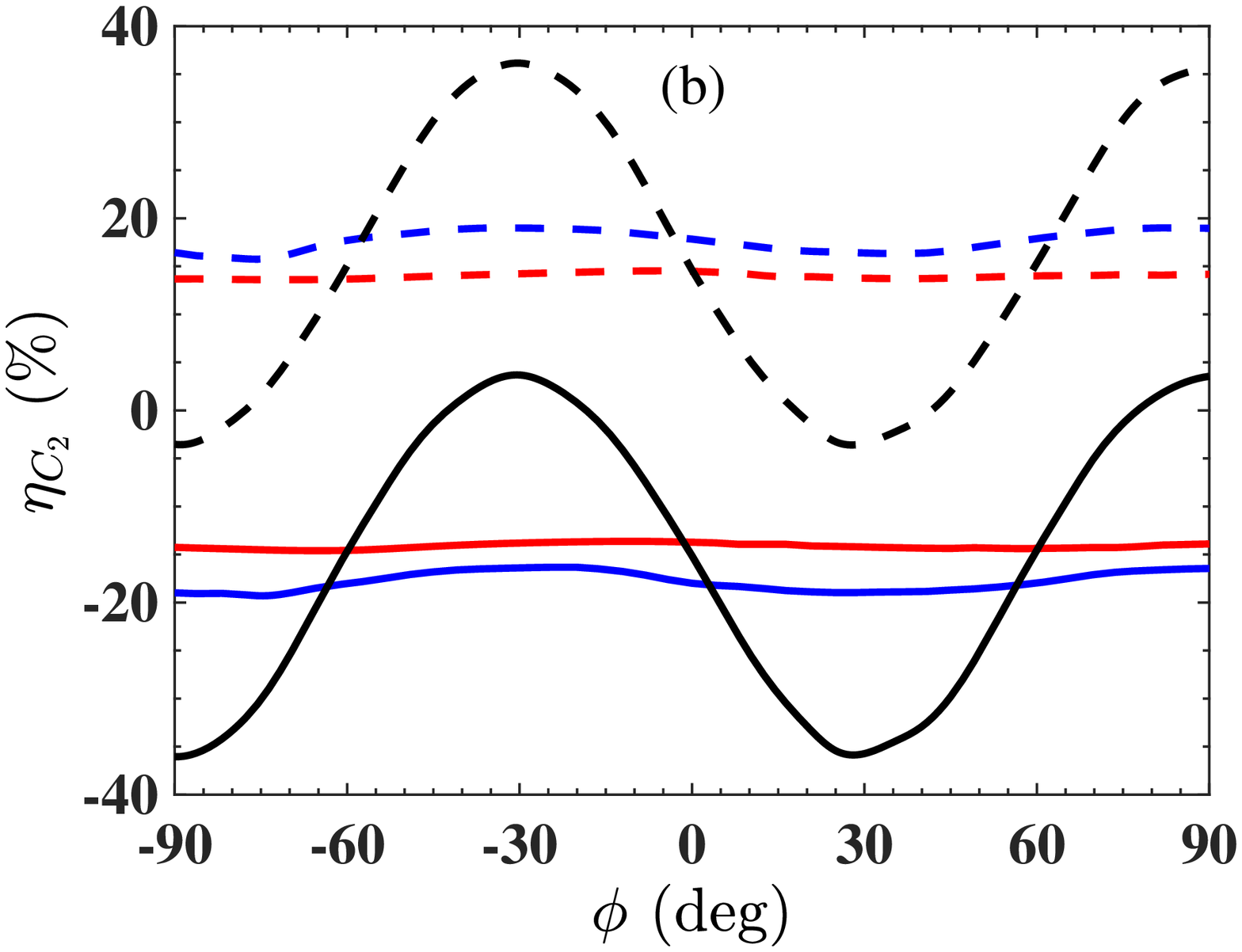}
		\end{tabular}
	\end{center}
\caption{Residual valley polarization generated by a circularly polarized pulse for (a) the first CB, $\eta_{C_{1}}$ (red lines) and (b) the second CB, $\eta_{C_{2}}$ (blue lines),   as a function of the in-plane orientation of the pulse.  Here,  $F_{0}=0.3~\mbox{V}/\mbox{\AA}$ (red lines),  $F_{0}=0.5~\mbox{V}/\mbox{\AA}$ (blue lines),  and  $F_{0}=0.7~\mbox{V}/\mbox{\AA}$ (black lines).  Further, for solid lines $\theta=80^{\circ}$ while for dashed lines $\theta=-80^{\circ}$ and rest of the parameters are same as in Fig. \ref{fig3}.}
\label{fig6}
\end{figure}

\begin{figure}[ht!]
	\begin{center}
		\begin{tabular}{cc}
			\includegraphics[scale=0.8]{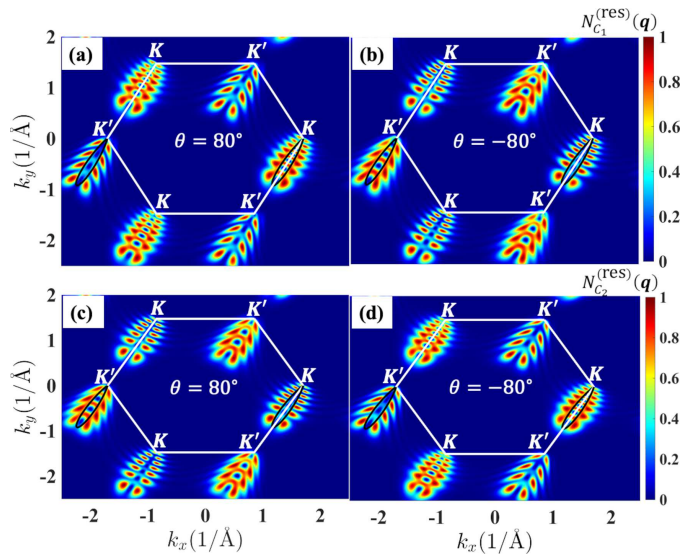}
		\end{tabular}
	\end{center}
\caption{Residual CB population distribution $N^{\mbox{(res)}}_{C_{2}}(\mathbf{k})$  for the first CB of the BLG for different values of angle of incidence ($\theta$) and angle of in-plane polarization ($\phi$). The boundary of the first Brillouin zone is shown by the white lines. Other parameters are the same as in Fig. \ref{fig3}.}
\label{fig7}
\end{figure}

As explained above, a finite angle of incidence of the applied pulse gives rise to its perpendicular component and to ellipticity as well. Such an ellipticity can further be manipulated by changing the in-plane orientation of the circularly polarized pulse which results in different CB population distributions around Dirac points thereby producing valley polarization in the system. This effect is illustrated in Fig. \ref{fig6} where the population imbalance of the $K$ and $K^{\prime}$-valleys for $C_{1}$ and $C_{2}$ is plotted as a function of in-plane orientation, $\phi$, for different field amplitudes and angles of incidence. For $F_{0}\le 0.5~\mbox{V}/\mbox{\AA}$, the valley polarization gets weakly influenced by the orientation of the circular pulse. However, for $F_{0}=0.7~\mbox{V}/\mbox{\AA}$ substantial contrast in the populations is achieved  at the Dirac points of $C_{1}$ and $C_{2}$. Clearly, large, up to 36 \%, 
valley polarization, $\eta_{C_{1}}$, is generated after the pulse for $\phi=-30^{\circ},90^{\circ}$ and $\theta=80^{\circ}$. However, the situation is reversed for $\eta_{C_{2}}$ where highest value of valley polarization is $-36\%$ for $\phi=30^{\circ},-90^{\circ}$. Further, the effect gets altered for $\theta=-80^{\circ}$, as shown in Fig. \ref{fig6}.  To quantify the influence of orientation of the circular pulse on the valley polarization, we plot residual CB populations in Fig. \ref{fig7} for $\phi=30^{\circ}$, $F_{0}=0.7~\mbox{V}/\mbox{\AA}$, and $\theta=\pm 80^{\circ}$. A clear strong contrast in the populations at $K$ and $K^{\prime}$ valleys is observable which is the reason for high valley polarization.

\subsection{Linearly polarized pulse followed by a circularly polarized pulse}

\begin{figure}[ht!]
	\begin{center}
		\begin{tabular}{c}
			\includegraphics[scale=0.45]{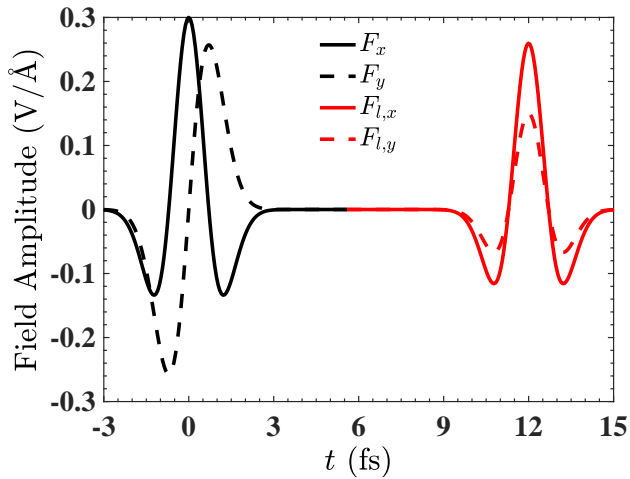}
		\end{tabular}
	\end{center}
	\caption{ Pulse sequence where a linearly polarized pulse (red lines) follows a circularly polarized pulse (black lines). The parameters chosen are $F_{l0}=0.3~\mbox{V}/\mbox{\AA}$, $\phi_{l}=30^{\circ}$, $\tau_{l}=1$ fs , $t_{0}=12$ fs and rest of the parameters are same as used in Fig. \ref{fig2}. }
	\label{fig8}
\end{figure}

As discussed in the previous section, a circularly polarized optical pulse with oblique incidence generates a valley polarization in BLG. The valley polarization depends on the angle of incidence,  the magnitude,  and in-plane orientation of the pulse.  Now, such a valley polarization generated by circular pulse can  also be  tuned by a \textit{normally} incident linearly polarized pulse, which is applied after the circularly polarized pulse [see Fig. \ref{fig8}]. 
To illustrate how a linearly polarized pulse affects the valley polarization, we fix the parameters of the circularly polarized 
pulse at  $\theta=\pm 80^{\circ}$, $\phi=0^{\circ}$, $F_{0}=0.5~\mbox{V}/\mbox{\AA}$, and change the parameters of the linearly 
polarized pulse, which are its amplitude, $F_{l0}$,  and angle of in-plane polarization, $\phi_{l}$.
The profile of the pulse is given by the following expression
\begin{equation}
F_{l,x}=F_{l}\cos\phi_{l}\label{eq9}
\end{equation}
\begin{equation}
F_{l,y}=F_{l}\sin\phi_{l}\label{eq10}
\end{equation}
where,
\begin{align}
F_{l}=F_{l0}\left(1-2u_{l}^{2}\right)e^{-u_{l}^{2}}\label{eq11}
\end{align}
Here $u_{l}=\left(t-t_{0}\right)/\tau_{l}$, $t_{0}$ is the center of the linearly polarized pulse and $\tau_{l}$ represents its linewidth which is equal to 1 fs in our case.  

\begin{figure}[ht!]
	\begin{center}
		\begin{tabular}{cc}
			\includegraphics[scale=0.75]{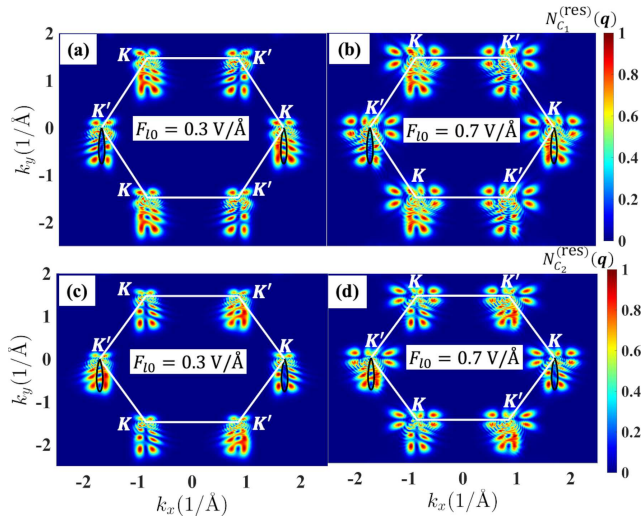}
		\end{tabular}
	\end{center}
\caption{Residual CB population distribution (a,b) $N^{\mbox{(res)}}_{C_{1}}(\mathbf{k})$  for the first CB and (c,d) $N^{\mbox{(res)}}_{C_{1}}(\mathbf{k})$  for the second CB of the BLG for different values of the amplitude of linearly polarized pulse ($F_{l0}$). Here, $F_{0}=0.5~\mbox{V}/\mbox{\AA}$, $\theta=80^{\circ}$, $\phi=0^{\circ}$ and $\phi_{l}=0^{\circ}$. The boundary of the first Brillouin zone is shown by the white lines and separatrix is represented by black lines. Other parameters are the same as in Fig. \ref{fig3}.}
\label{fig9}
\end{figure}

In Fig. \ref{fig9}, we show how a linearly polarized pulse manipulates the residual CB population generated by a circularly polarized pulse. As is evident from Fig. \ref{fig9}, that linearly polarized pulse produces fringes which interfere with textures generated by circular pulse and the effect becomes pronounced with increase in the amplitude of the successive pulse. Thus, residual CB population and hence the valley polarization can be controlled by means of a linearly polarized pulse.

\begin{figure}[ht!]
	\begin{center}
		\begin{tabular}{c}
			\includegraphics[scale=0.45]{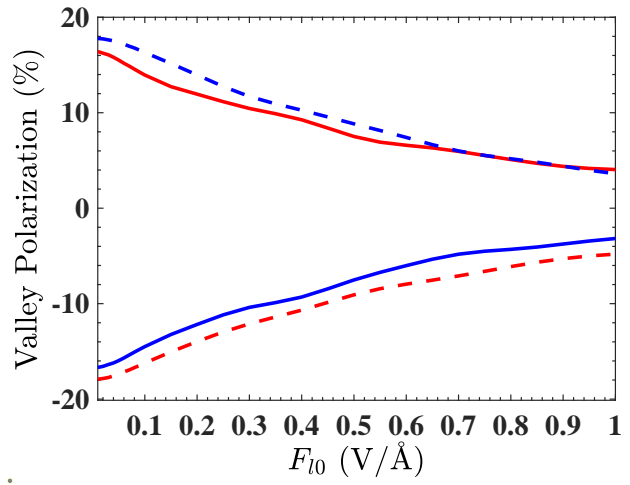}
		\end{tabular}
	\end{center}
	\caption{Valley polarization after linearly polarized pulse for the first CB, $\eta_{C_{1}}$ (red lines),   and the second CB, $\eta_{C_{2}}$ (blue lines),   as a function of the amplitude of the pulse, $F_{l0}$ for $\theta_{l}=0^{\circ}$ .  The parameters of the circularly polarized pulse are  $F_{0}=0.5~\mbox{V}/\mbox{\AA}$, $\phi=0^{\circ}$, $\theta=80^{\circ}$ (solid lines),  $\theta=-80^{\circ}$ (dashed lines), and the other parameters are same as in Fig. \ref{fig2}.}
	\label{fig10}
\end{figure}

To quantify the effect of linearly polarized pulse, in Fig. \ref{fig10},  we show the residual valley polarization for CBs 1 and 2  as a function of the amplitude of the linearly polarized 
pulse for different angles of incidence of the circular pulse. The valley polarization shows strong sensitivity to the amplitude of the linearly polarized pulse. Both $\eta_{C_{1}}$ and $\eta_{C_{2}}$ decrease with amplitude $F_{l0}$. Thus, the linearly polarized pulse depolarizes the system. For $\theta=\pm80^{\circ}$, the valley polarization, $\eta_{C_{1}}$ and $\eta_{C_{2}}$ changes by almost 13\% when the amplitude of the linearly polarized pulse increases from its small values to 1 V/\AA. This variation strongly depends on the angle of incidence of the circular pulse. 

\begin{figure}[ht!]
	\begin{center}
		\begin{tabular}{c}
			\includegraphics[scale=0.45]{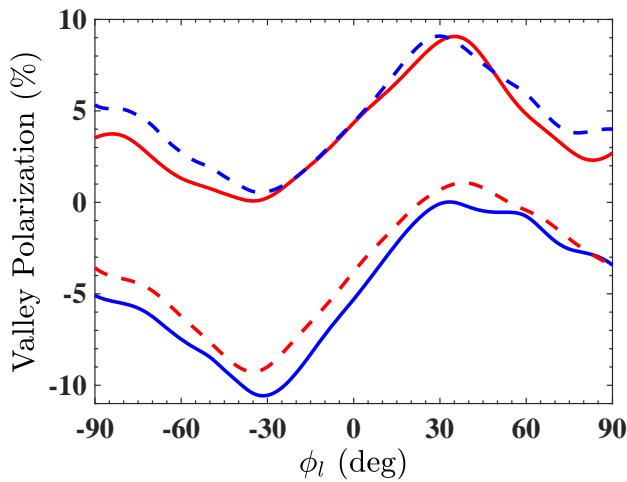} 
		\end{tabular}
	\end{center}
	\caption{Valley polarization after linearly polarized pulse for the first CB, $\eta_{C_{1}}$ (red lines),   and the second CB, $\eta_{C_{2}}$ (blue lines),   as a function of the angle of polarization $\phi_{l}$ for $F_{l0}=0.9~\mbox{V}/\mbox{\AA}$ .   Here, $\theta=80^{\circ}$ and $-80^{\circ}$ for solid and dashed lines, respectively.  Other parameters are same as in Fig. \ref{fig10}.}
	\label{fig11}
\end{figure}

 Above we have described that the valleys get depolarized with increase in the amplitude of the linearly polarized pulse. However, for a fixed field amplitude, it is further possible to tune the valley polarization by changing the angle of polarization of the linearly polarized pulse.  In Fig. \ref{fig11}, we show the valley polarization as a function of the angle of polarization,  $\phi_{l}$, for $F_{l0}=0.9~\mbox{V}/\mbox{\AA}$ and $\theta=\pm 80^{\circ}$. As is clear from Fig. \ref{fig11} that by selecting an appropriate in plane orientation of the linearly polarized pulse it is possible to depolarize as well as further polarize the system.  For instance, for $\theta=80^{\circ}$ and $0^{\circ}<\phi_{l}\le 35^{\circ}$,   valley polarization, $\eta_{C_{1}}$,  increases upto $5\%$  and for $35^{\circ}<\phi_{l}<90^{\circ}$ its depolarization is $\sim$ $7\%$.  Contrary to this,  $\eta_{C_{2}}$ depolarizes upto $5\%$ for $0^{\circ}<\phi_{l}\le 35^{\circ}$ and then it again attains a polarization of $\sim$ 4\%  for $35^{\circ}<\phi_{l}<90^{\circ}$. The situation gets reversed for negative angles of polarization of the linearly polarized pulse as shown in Fig. \ref{fig11}. Further, the angle of incidence of the circular pulse strongly influence the valley polarization.  Thus the angle of polarization of a linearly polarized pulse plays an important role to tune the valley polarization generated by a circularly polarized pulse.

\section{Conclusions}
\label{conclusion}

We have shown numerically that, in AB-stacked BLG, it is possible to generate the valley polarization by using a circularly polarized optical pulse of a single oscillation at an oblique incidence. Due to normal component of the optical field, such pulse 
dynamically breaks the inversion symmetry of BLG and opens a dynamical band gap, which finally affects the  
residual valley polarization of the system. The valley polarization can be tuned by the angle of incidence,  the amplitude, ans the angle of in-plane polarization of the circularly polarized pulse. For example, we have shown that at the angle of  incidence of $80^{\circ}$, in-plane polarization of $30^{\circ}$ and the field amplitude of 0.7 $\mbox{V}/\mbox{\AA}$ the valley polarization is around 36\%.

The valley polarization induced by a circularly polarized can be also controlled by a subsequent normally incident linearly polarized pulse.  The magnitude of the change of the valley polarization depends on the amplitude, and angle of in-plane polarization of the linearly polarized pulse,. Generally, the linearly polarized pulse depolarizes the system.  We have shown that the depolarization of up to 13$\%$ can be achieved by using strong linearly polarized pulses. Thus, while the circularly polarized pulse can be used to generate large valley polarization, the linearly polarized pulse, on the other hand, can be used to depolarize the system. 
   
\begin{acknowledgements}
	Major funding was provided by Grant No. DE-FG02-11ER46789 from the Materials Sciences and Engineering Division of the Office of the Basic Energy Sciences, Office of Science, U.S. Department of Energy. Numerical simulations have been performed using support by Grant Nos. DE-FG02-01ER15213 and DE-SC0007043 from the Chemical Sciences, Biosciences and Geosciences Division, Office of Basic Energy Sciences, Office of Science, US Department of Energy.  Supplementary funding came from (i) Grant No. N000-14-17-1-2588 from the Office of Naval Research (ONR), (ii) Grant No. FA9550-15-1-0037 from University of Central Florida, subcontracted by the Air Force Office of Scientific Research (AFOSR), and (iii) Grant No.  EFMA-1741691 from Emory University, subcontracted by the National Science Foundation (NSF). PK has been supported in part by the Grant  No. CMMI 1661618 from the National Science Foundation. We are indebted to late Prof. Mark I. Stockman for fruitful discussions.
\end{acknowledgements}

\end{document}